\documentclass[12pt,final,notitlepage,oneside,onecolumn,nobibnotes,
    nofootinbib, superscriptaddress,noshowpacs,centertags]{revtex4}

\begin{document}
%


\title{
Continuous Transitions between Discontinuous
Magnetohydrodynamic Flows of Plasma and Its Heating}

\author{\copyright\, 2013 ã. \quad
        \firstname{L.~S.}~\surname{Ledentsov} }
        \email{koob@mail.ru}
        \affiliation{Sternberg Astronomical Institute, Universitetskii pr. 13, Moscow, 119992 Russia \\ Moscow State University, Moscow, 119991 Russia}

\author{\firstname{B.~V.}~\surname{Somov} }
        \email{somov@sai.msu.ru}
        \affiliation{Sternberg Astronomical Institute, Universitetskii pr. 13, Moscow, 119992 Russia \\ Moscow State University, Moscow, 119991 Russia}

%
%
\begin{abstract}
The possibility that the type of discontinuous flow changes as the conditions gradually (continuously) change is investigated in connection with the problems arising when the results of numerical simulations of magnetic reconnection in plasma are interpreted. The conservation laws at a discontinuity surface in magnetohydrodynamics admit such transitions, but the socalled transition solutions for the boundary conditions that simultaneously satisfy two types of discontinuities should exist in this case. The specific form of such solutions has been found, and a generalized scheme of permitted transitions has been constructed on their basis. An expression for the jump in internal energy at discontinuity is derived. The dependence of the plasma heating efficiency on the type of discontinuity is considered.\end{abstract}
%
%


\maketitle


%
%
\section{INTRODUCTION}
   \label{sec1}

Discontinuous plasma flows in a magnetic field are present in various kinds of technical facilities and devices of practical significance
\cite{Sutton-65,Lukianov-75,Morozov-06},
%
in laboratory and numerical experiments
\cite{Biskamp-97,Imshennik-97,Buchner-03},
%
and in astrophysical conditions—especially in connection with the magnetic reconnection effect
\cite{Syrovatskii-62,Petschek-64,Brushlinskii-80,Hones-84,Hoshino-01}.
%
The question about plasma heating to very high temperatures is generally important in this case 
\cite{Orta-03}.
For example, as applied to solar flares, the appearance of plasma in the solar corona whose electron temperature exceeds considerably 10 keV 
\cite{Sui-03}
needs to be explained.
%
%

Presentday numerical experiments simulating the reconnection process, for example, in the approximation of dissipative magnetohydrodynamics (MHD), demonstrate a slightly smoothed picture of discontinuous flows in the vicinity of a reconnecting current layer
\cite{Shimizu-03,Ugai-05,Ugai-08,Zenitani-11}.
%
When the results of such calculations, especially threedimensional ones, are interpreted, it is difficult to unequivocally identify the type of a particular discontinuity by some incomplete and averaged set of attributes. If this difficulty has been successfully overcome, then a second problem arises; it consists in explaining the simultaneous presence of different types of discontinuities gradually transforming into one another in such experiments. In this case, the task of interpreting the picture of transitions is often complicated by the appearance of nonevolutionary discontinuities
\cite{Bezrodnykh-11}.
%
%

rom a theoretical point of view, the following facts play a fundamental role in investigating the properties of discontinuous plasma flows. The equations of
ordinary hydrodynamics are known to have only two types of discontinuous solutions: a tangential discontinuity and a shock. In magnetohydrodynamics (MHD), the presence of a magnetic field in plasma leads to the existence of fast, slow, Alfv'en shocks and other discontinuities
\cite{Syrovatskii-57,Anderson-63}.
Moreover, in contrast to hydrodynamics, continuous transitions 
\cite{Syrovatskii-56,Polovin-60}
between different types of discontinuous solutions as the flow conditions change continuously are possible in MHD. This occurs through the socalled transition solutions that simultaneously satisfy the conditions of two types of discontinuities.
%
%

The first description of transition solutions
\cite{Syrovatskii-56}
contained only four types of discontinuous flows: a tangential discontinuity~($ T $) and Alfv'en~($ A $), oblique~($ S $) and perpendicular~($ S_{\bot} $) shocks. The corresponding scheme of continuous transitions between discontinuous solutions of the equations of ideal MHD showed such transitions to be possible in principle, but it was definitely incomplete. First, it did not have some of the discontinuous solutions, in particular, the parallel shock~($ S_{\, \|} $) and the contact discontinuity~($ C $).
Second, the block of oblique shocks ($ S $) combined several different discontinuities at once: fast~($ S_+ $) and slow~($ S_- $) shocks, switchon~($ S_{\, \rm on} $)
and switchoff ~($ S_{\, \rm off} $),
shocks, and trans-Alfv'en ($ Tr $), shocks, the possibility of transitions between which requires a separate consideration. Subsequently, this picture of transitions was supplemented based on the correspondence between shocks and smallamplitude waves
\cite{Somov-94}.
%
Although this approach allows the possible transitions and even their conditions to be correctly specified, it provides no
description of the specific form of transition solutions between the discontinuities under consideration.
%
%

An abrupt change, a jump in plasma parameters and magnetic field, occurs at a discontinuity surface. The type of discontinuous MHD solution, i.e., its character, is determined by the changes in the plasma density, flow velocity, and magnetic field frozen into the plasma. In addition, plasma heating occurs at the discontinuity surface, whose magnitude, of course, also depends on the type of discontinuity but does not determine its classification attributes: the density continuity or jump and the presence or absence of perpendicular velocity~$ v_{\bot} $ and magnetic field~$ B_{\bot} $.

This paper is devoted to analyzing the boundary conditions at a discontinuity surface that are derived from the equations of ideal MHD and that constitute a system of nonlinear eighthorder algebraic equations. Depending on the choice of parameters, the solutions of this system describe a discontinuity of a particular type. The paper is structured as follows. Initially, we seek for the transition solutions at which the type of discontinuity changes, which allows us to systematize such solutions into a generalized scheme of permitted transitions. This can be done without invoking the boundary condition corresponding to the law of conservation of the energy flux through the discontinuity surface. Subsequently, using the energy conservation law, we derive the equation that describes the jump in plasma internal energy at the discontinuity surface and consider the plasma heating efficiency depending on the type of discontinuity.
%
%

\vspace{2mm}

%
%
\section{BOUNDARY CONDITIONS AT DISCONTINUITY}
\label{sec2}

At the surface of an MHD discontinuity, the plasma density, pressure, flow velocity, and the magnetic field direction and strength can change abruptly at a distance comparable to the particle mean free path. The physical processes inside such a jump are determined by kinetic phenomena in plasma, both laminar and turbulent ones
\cite{Longmire-66,Tideman-71}.
%
In the approximation of dissipative MHD, the internal structure of a discontinuous flow is defined by dissipative transport coefficients (the viscosity and electric conductivity) and the thermal conductivity
\cite{Sirotina-60,Zeldovich-66}.
However, in the approximation of ideal MHD, the jump has zero thickness, i.e., it occurs at some discontinuity surface.
%
%

We will consider a plane discontinuity surface, which is appropriate for areas of a sufficiently small size compared to the radius of curvature of the discontinuity surface. Let us introduce a Cartesian coordinate system in which the observer moves with the discontinuity surface located in the $ (y,z) $,
plane in the direction of the~$ x $.
axis. In the approximation of ideal MHD, we neglect the plasma viscosity, thermal conductivity, and electric resistivity. The boundary conditions for the MHD equations at discontinuity then take the form of the following conservation laws (see
\cite{Syrovatskii-57}):
%
\begin{equation}
    \left\{ \, B_{x} \, \right\} = 0 \, ,
    \label{GRx1}
\end{equation}
\begin{equation}
    \left\{ \, \rho \, v_{x} \, \right\} = 0 \, ,
    \label{GRx2}
\end{equation}
\begin{equation}
    \left\{ \, v_{x} B_{y} - v_{y} B_{x} \, \right\} = 0 \, ,
    \label{GRx3}
\end{equation}
\begin{equation}
    \left\{ \, v_{x} B_{z} - v_{z} B_{x} \, \right\} = 0 \, ,
    \label{GRx4}
\end{equation}
\begin{equation}
    \left\{ \,
    \rho \, v_{x} v_{y} - \frac{1}{4 \pi} \, B_{x} B_{y} \,
    \right\} = 0 \, ,
    \label{GRx5}
\end{equation}
\begin{equation}
    \left\{ \,
    \rho \, v_{x} v_{z} - \frac{1}{4 \pi} \, B_{x} B_{z} \,
    \right\} = 0 \, ,
    \label{GRx6}
\end{equation}
\begin{equation}
    \left\{ \, p + \rho \, v^{\,2}_{x}
  + \frac{ B^{\, 2} }{ 8 \pi } \,
    \right\}
  = 0 \, ,
    \label{GRx7}
\end{equation}
\begin{equation}
    \left\{ \, \rho \, v_{x} \left( \frac{v^{\,2}}{2} + \epsilon
  + \frac {p}{\rho} \right)
  + \frac{1}{4 \pi} \left( \, B^{\,2} v_{x}
  - ( \, {\bf v} \cdot {\bf B} \, ) \, B_{x} \, \right) \right\}
  = 0 \, .
    \label{GRx8}
\end{equation}

\vspace{1mm}
\noindent
Here, the curly brackets denote the difference between the values of the quantity contained within the brackets on both sides of the discontinuity plane. For example, Eq.~(\ref{GRx1}) implies the continuity of the normal magnetic field component:
$$ \left\{ \, B_{x} \, \right\} =  B_{x2} - B_{x1} = 0 \, . $$
The quantities marked by the subscripts ``1'' and ``2'' refer to the side corresponding to the plasma inflow and outflow, respectively.
%
%

In contrast to the boundary conditions in ordinary hydrodynamics, the system of boundary conditions (\ref{GRx1})--(\ref{GRx8})
does not break up into a set of mutually exclusive groups of equations and, hence, in principle it admits continuous transitions between different types of discontinuous solutions as the plasma flow conditions change continuously. Since a smooth transition between discontinuities of various types is possible, the local external attributes of the flow near the discontinuity plane are taken as a basis for their classification: the presence or absence of a mass flux and a magnetic flux through the discontinuity, the density continuity or jump.
%
%

For plasma flows in the $ ( x, y ) $
plane, the formula relating the inclination angles of the magnetic field vector to the normal of the discontinuity surface
$ x = 0 $ to the densities $ \rho_1 $ and $ \rho_2 $,
the magnetic field component $ B_x $ and the mass flux
$ m = \rho v_x \,$ follows from the first seven equations of system
(\ref{GRx1})--(\ref{GRx8})
\cite{Ledentsov-11}:
\begin{displaymath}
    {\rm tan} \, \theta_2
  = {
    2 \,
    \left( \, { B_x^{\, 2} } / \, { 4 \pi } - m^{\,2} \,
      {\tilde r} \,
    \right) + m^{\,2}\, \{\, r\, \} \over
    2 \,
    \left( \, { B_x^{\, 2} } / \, { 4 \pi } - m^{\,2} \,
      {\tilde r} \,
    \right) - m^{\,2}\, \{\, r\, \}
    } \,\,
    {\rm tan} \, \theta_1 \, .
   \label{GRx9}
\end{displaymath}
Here, $ {\rm tan} \, \theta  = B_y / \,B_x $, $ r = 1  / \rho \, $,
and the tilde marks the mean values of the quantities, $ \tilde r = (\, r_1 + r_2 \,) / \, 2 $.
Rewriting this formula by expanding the jumps $ \{ r \} $ and the means
 $ \tilde r $, we obtain
\begin{displaymath}
    {\rm tan} \, \theta_2
  = {m^{\,2} \cdot 4 \pi r_1 / B_x^{\, 2} - 1
    \over
    m^{\,2} \cdot 4 \pi r_2 / B_x^{\, 2} - 1}
    \,\, {\rm tan} \, \theta_1 \, .
   \label{GRx9.1}
\end{displaymath}
Denote
$ m_{\rm off}^{\,2} = B_x^{\, 2} / 4 \pi r_1 $ and
$ m_{\rm on}^{\,2}  = B_x^{\, 2} / 4 \pi r_{2 \,} $;
as will be shown below, $ m_{\rm off}$ and $m_{\rm on} $ correspond to the mass flux through the switchoff and switchon shocks. Note that
$ m_{\rm off}~\leq~m_{\rm on} $, because, in view of Zemplen's theorem (see  \cite{Landau-82}, Section 72),
$ r_2 \leq r_1 $.
The formula for the field inclination angles takes the form
\begin{equation}
   {\rm tan} \, \theta_2
 = \frac{ m^2 /m_{\rm off}^2 - 1 }{ m^2 / m_{\rm on}^2 - 1} \,
   {\rm tan} \, \theta_1 \, .
   \label{GRx9.2}
\end{equation}

A constraint on the possible mass flux through the discontinuity follows from the condition for the existence of nontrivial solutions to the system of seven equations under consideration
\cite{Ledentsov-11}:
\begin{equation}
    m^{\,2} < { B_x^{\, 2} \over 4 \pi {\tilde r} } \, ,
    \label{GRx10}
\end{equation}
or
\begin{equation}
    m^{\,2} > { B_x^{\, 2} + \tilde{B}_y^{\, 2}
    \over 4 \pi {\tilde r} } \, .
    \label{GRx11}
\end{equation}
%
%
Denote
$ m_{_{\rm A}}^{\,2} = B_x^{\, 2} / 4 \pi {\tilde r} $ and
$ m^{\,2}_\bot = \tilde{B}_y^{\, 2} / \, 4 \pi {\tilde r} $.
The quantity $ m_{\rm A} $ corresponds to the Alfv'en mass flux. Since
$ r_2 \leq \tilde r \leq r_1 $, òî
$ m_{\rm off} \leq~m_{_{\rm A}} \leq m_{\rm on} $.
We will write conditions (\ref{GRx10}) and (\ref{GRx11}) as:
\begin{equation}
    m^{\,2} < m_{_{\rm A}}^{\, 2} \, ,
    \label{GRx10.1}
\end{equation}
\begin{equation}
    m^{\,2} > m_{_{\rm A}}^{\,2} + m_{\bot}^{\,2} \, .
    \label{GRx11.1}
\end{equation}
%
%
Note that
$ m_{\rm off} $, $ m_{\rm A} $ and $ m_{\rm on} $ are not independent but
are related by the relation
\begin{equation}
    m_{_{\rm A}}^{\, 2}
  = \frac{ 2 \,\, m_{\rm on}^{\, 2} \, m_{\rm off}^{\, 2} }
         { m_{\rm on}^{\, 2} + m_{\rm off}^{\, 2} } \, .
    \label{ma}
\end{equation}
This can be easily verified by expanding the mean~$ \tilde r $ in the definition of $ m_{\rm A}^{\, 2} $.
%
%

Based on these results, below we will consider the properties of discontinuous flows; more specifically, we will establish the possible transitions between them.
%
%

\vspace{1mm}

%
%
\section{TWODIMENSIONAL DISCONTINUOUS FLOWS}
   \label{sec3}

We will begin seeking transition solutions with a search for the conditions of possible transitions between various types of twodimensional MHD flows
($ v_z = 0 $, $ B_z = 0 $), i.e., flows for which the velocity field and the magnetic field lie in the
$ (x, y) $.
plane. We call such discontinuous flows plane or twodimensional ones. Then, we will find the transition solutions that correspond to them and establish the form of the solutions that are the transition ones to threedimensional discontinuous flows.
%
%

\begin{figure} 
    \centerline{\includegraphics[width=150mm]{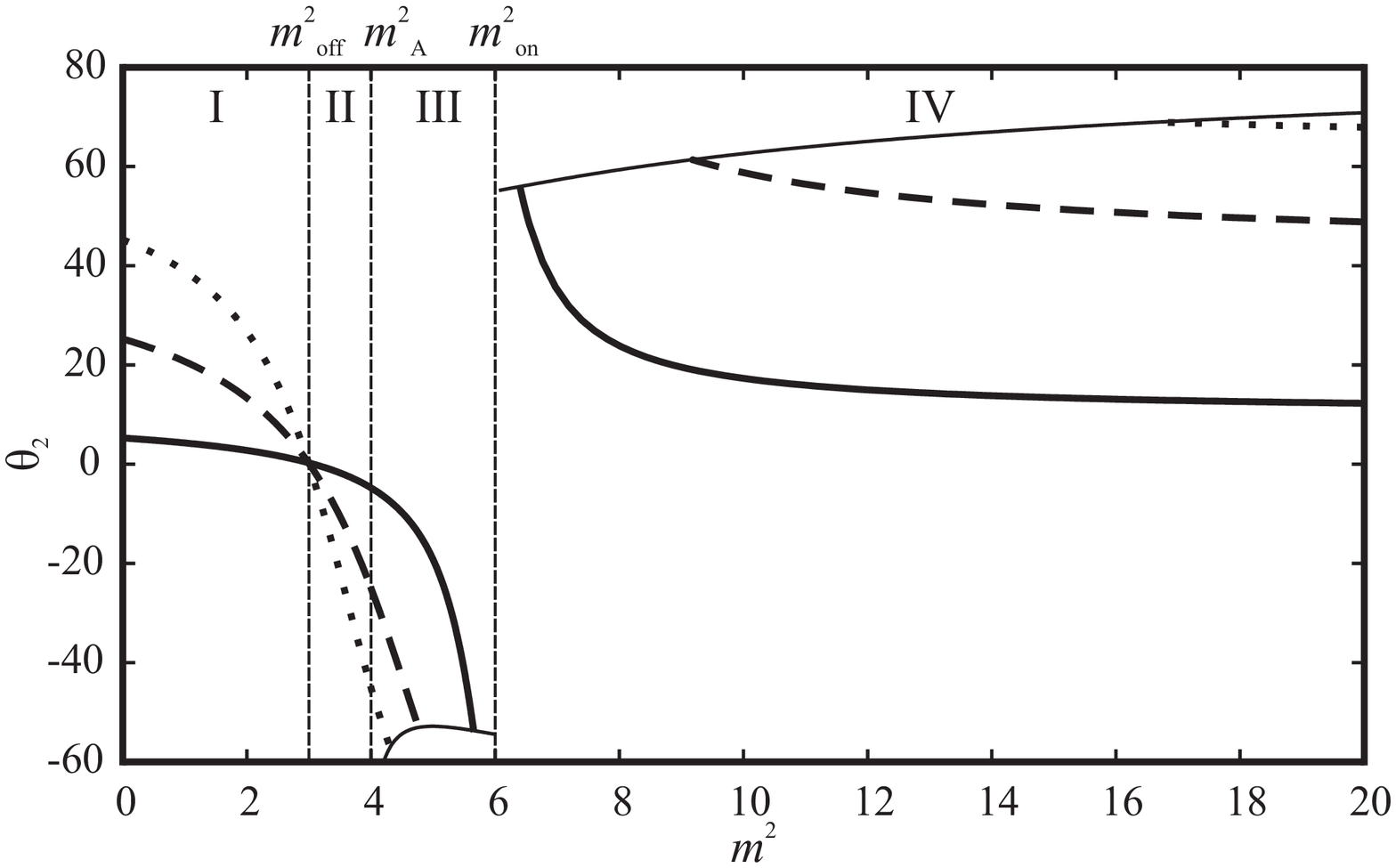}}
    \vspace*{3cm}
\caption{\rm Inclination angle $ \theta_{2} $ of the magnetic field behind the discontinuity plane versus square of the mass flux
       $ m^2 $ for various angles $ \theta_{1} $
       $ 5^{\circ} $ (solid lines),
       $ 25^{\circ} $ (dashed lines), and
       $ 45^{\circ} $ (dotted lines).}
    \label{fig1}
\end{figure}

Equation (\ref{GRx9.2}) along with conditions
(\ref{GRx10.1})--(\ref{GRx11.1})
describes the dependence of the magnetic field inclination angles on the mass flux through the discontinuity. This dependence can be specified either by the two parameters
$ m_{\rm off} $ and $ m_{\rm on} $,
or, for example, by the quantities $ \rho_1 $ and $ \{\rho\} $.
Since we are interested in the classification attributes of discontinuities (i.e., the qualitative changes of the relation between the angles $ \theta_1 $ and $ \theta_2 $ when varying $ m^2 $),
for the time being, we will consider (\ref{GRx9.2}) without any specific application to certain physical conditions in plasma. We will choose the parameters
from clarity considerations. Let
$ m_{\rm off}^{\,2} $, $ m_{_{\rm A}}^{\,2} $ and
$ m_{\rm on}^{\,2} $ be offA on
related as $ 3 : 4 : 6 $.
We will measure the square of the mass fluxin units of 
$ m_{_{\rm A}}^{\,2} / 4 $;
then,
$ m_{\rm off}^{\,2} = 3 $
and
$ m_{\rm on}^{\,2} = 6 $.
%
%
%
The dependence $ \theta_2 ( m^2 ) $ is shown in Fig. \ref{fig1}
%
%
or three
values of the angle $ \theta_1 $, The corresponding curves behave identically. First, they intersect at one point at
$ m^2 = m_{\rm off}^{\,2} $, $ \theta_2 = 0 $ here.Second,
$ \theta_2 \rightarrow - \, \theta_1 $ when
$ m^2 \rightarrow m_{_{\rm A}}^{\,2} = 4 $.
for each curve. Third, they all have a region that does not satisfy conditions (\ref{GRx10.1}) and (\ref{GRx11.1}), located near
$ m^2 = m_{\rm on}^{\,2} $.
%
%

Let us separate out the regions in Fig.~\ref{fig1}
%
%
each of which is characterized by its own behavior of the
dependenceof $ \theta_2 $ on $ m^{2} $.
Inregion $ {\rm I} $ ($ 0 < m^2 < m_{\rm off}^{\,2} $), the
tangential component $ B_{y2} $
of the magnetic field vector $ {\bf B}_{2} $ behind the discontinuity surface decreases with
increasing $ m^{2} $.
In this case, $ 0 < \theta_2 < \theta_1 $, i.e., when passing through the discontinuity surface, the tangential field component weakens but remains positive. At $ m^{2} = m_{\rm off}^{\,2} $ when crossing the discontinuity plane,
$ B_{y2} $  becomes zero. In region~$ {\rm II} $
($ m_{\rm off}^{\,2} < m^2 < m_{_{\rm A}}^{\,2} $)
 $ B_{y2} $
is negative and increases in magnitude, but
$ - \theta_1 < \theta_2 < 0 $.
In region~$ { \rm III } $
($ m_{_{\rm A}}^{\,2} < m^2 < m_{\rm on}^{\,2} $), just as in region~$ {\rm II} $, $ B_{y} $  changes its sign when crossing the discontinuity plane. Now, however,
$ B_{y} $ increases in magnitude
($ \theta_2 < - \theta_1 $), remaining negative. Finally, in region~$ {\rm IV} $ ($ m^2 > m_{\rm on}^{\,2} $)
the magnetic field is amplified ($ \theta_2 > \theta_1 $) with its tangential component retaining the positive sign.
%
%

Consider the behavior of the function $ \theta_2 ( m^2, \theta_1 ) $
near the boundary of regions  $ {\rm II} $ and $ {\rm III} $, where
$ m^{\,2} = m_{_{\rm A}}^{\,2} $.
The domain of definition of the function $ \theta_2 ( m^2, \theta_1 ) $
to the left and the right of $ m_{_{\rm A}}^{\,2} $ is specified by conditions (\ref{GRx10.1}), and (\ref{GRx11.1}), respectively. In view of (\ref{GRx9.2}) and (\ref{ma}), $ {\rm tan} \, \theta_2 \to - {\rm tan} \, \theta_1 $ when $ m^{\,2} \to m_{_{\rm A}}^{\,2} $,
i.e., $ \tilde{B}_y \to 0 $.
Inequality (\ref{GRx11.1}) transforms to
$ m^{\,2} > m_{_{\rm A}}^{\,2} $.
in this case. Therefore, the function $ \theta_2 (m^2, \theta_1) $ in both region
$ {\rm II} $ and region $ {\rm III} $
is defined near $ m_{_{\rm A}}^{\,2} $.
However, the right part of condition $ m^2 $ also increases with (\ref{GRx11.1}).
The equality 
$ m^2 = m_{_{\rm A}}^{\,2} + m_{\bot}^{\,2} $ is established at some value of $ m^2 $ in region $ {\rm III} $
and the strongest trans-Alfv'en shock (increasing the magnetic energy to the greatest extent)
takes place. As the mass flux increases further, $ m^2 $ can
not satisfy conditions (\ref{GRx10.1}) and (\ref{GRx11.1})
until $ m^2 $ again
becomes equal to
$ m_{_{\rm A}}^{\,2} + m_{\bot}^{\,2} $.
This occurs in region $ {\rm IV} $, where the strongest fast shock is observed.
%
%

Let us derive the equation of the curve bounding the function $ \theta_2 ( m^2, \theta_1 ) $,
and, hence, the strongest (for given plasma parameters) fast and trans-Alfv'en shocks. Setting $ m^{\,2} $ equal to the right part of condition (\ref{GRx11}),
we find
\begin{displaymath}
    B_{y1}
  = \pm \, 2 \,
    \sqrt{4 \pi \tilde r \, m^2 - B_x^2} \, - B_{y2} \, ,
    \label{max2}
\end{displaymath}
where the plus and minus correspond to regions IV and III, respectively. Dividing the derived equation by $ B_x $, we have
\begin{equation}
   {\rm tan} \, \theta_1
 = \pm \, 2 \,
   \sqrt{ m^2 / \,m_{_{\rm A}}^{\, 2} - 1} \,
 - {\rm tan} \, \theta_2 \, .
   \label{max3}
\end{equation}
Substituting Eq. (\ref{max3}) into (\ref{GRx9.2}), we obtain the equation of the soughtfor curve
\begin{displaymath}
   {\rm tan} \, \theta_2 =
   \pm \,
   \frac{ 2 \, m^2 / \, m_{\rm off}^2 - 2 }{
               m^2 / \, m_{\rm off}^2 + \,
               m^2 / \, m_{\rm on}^2 - 2 } \,\,
   \sqrt{\frac{m^2}{m_{\rm A}^2} - 1} \, .
\end{displaymath}
Let us simplify it by using relation~(\ref{ma}).
We have
\begin{equation}
   {\rm tan} \, \theta_2 =
   \pm \,
   \frac{ m^{\,2} / \, m_{\rm off}^{\,2} - 1}{
   \sqrt{ m^{\,2} / \,m_{\rm A}^{\,2} - 1}} \, .
   \label{max}
\end{equation}
The corresponding curves are represented in Fig.~\ref{fig1}
%
%
by the thin lines.
%
%

Thus, we have shown precisely how the behavior of the relation between the magnetic field inclination angles and, consequently, the type of MHD discontinuity changes with increasing mass flux. Regions I and II correspond to the slow shocks that, respectively, do
not reverse~($ S_{-}^{\,\uparrow} $) and reverse~($ {S_{-}^{\,\downarrow}} $) the tangential field ––
component. Regions III and IV correspond to the trans-Alfv'en~($ {Tr} $), and fast~($ {S_+} $).
shocks, respectively. In this case, transition solutions for the discontinuities corresponding to the adjacent regions are realized at the mass flux demarcating these regions.
%
%

Now, we will seek transition solutions in order of increasing mass flux $ m $, starting from $ m = 0 $.
Consider the transition between the contact discontinuity ($ {C} $)
at $ m^2 = 0 $
and the slow shock in region I. The boundary conditions for twodimensional discontinuities follow from
(\ref{GRx1})--(\ref{GRx7})
when $ v_z = 0 $ and $ B_z = 0 $ are substituted:
\begin{equation}
    \left\{ \, B_{x} \, \right\} = 0 \, ,
    \quad
    \left\{ \, \rho \, v_{x} \, \right\} = 0 \, ,
    \quad
    \left\{ \,
    p + \rho \, v^{\,2}_{x} + \frac{ B_{y}^{\,2} }{ 8 \pi } \,
    \right\}
  = 0 \, ,
    \label{GRx12}
\end{equation}
\begin{displaymath}
    \left\{ \, v_{x} B_{y} - v_{y} B_{x} \, \right\} = 0 \, ,
    \quad
    \left\{ \,
    \rho \, v_{x} v_{y} - \frac{ 1 }{ 4 \pi } \, B_{x} B_{y} \,
    \right\} = 0 \, .
\end{displaymath}

\vspace{0.3mm}

\noindent
The solution of these equations in region I presented in Fig.~\ref{fig1}
%
%
corresponds to the slow shock
($ S_{-}^{\,\uparrow} $),
that does
not change the sign of the tangential magnetic field component. This can be easily verified at small $ m^2 $.
It
follows from Eq.~(\ref{GRx9.2}), which, of course, remains applicable under conditions of twodimensional discontinuous flows, that
\begin{displaymath}
   {\rm tan} \, \theta_2 \approx
   \left( 1 -
   \frac{ m^2 }{ m_{\rm off}^2 }
 + \frac{ m^2 }{ m_{\rm on}^2  } \right) \,
   {\rm tan} \, \theta_1 \, ,
\end{displaymath}
i.e., $ 0 < \theta_2 < \theta_1 $, as it must be in the slow shock. Moreover, when $ m^2 \rightarrow 0 $, i.e., 
$ v_x \rightarrow 0 $, it follows from (\ref{GRx12}) that
$ \left\{ \, v_{y} \, \right\} \rightarrow 0 $ and
$ \left\{ \, B_{y} \, \right\} \rightarrow 0 $,which is the only limiting case for the slow shock.


%
%

It remains to show that when $ v_x \rightarrow 0 $ conditions
(\ref{GRx12}) transform to the boundary conditions at the contact discontinuity. Indeed, substituting $ v_x = 0 $ into
(\ref{GRx12}) gives
%
\begin{equation}
    \left\{ \, B_{x} \, \right\} =
    \left\{ \, v_{y} \, \right\} =
    \left\{ \, B_{y} \, \right\} =
    \left\{ \, p \, \right\} = 0 \, ,
        \label{GRx17}
\end{equation}
At the contact discontinuity, the jump in density $ \{ \rho \} $
is
nonzero. Otherwise, all quantities remain continuous. Thus, solution (\ref{GRx17})
simultaneously describes both the slow shock in the limit $ v_x \rightarrow 0 $,and the contact discontinuity, i.e., it is the corresponding transition solution.
%
%

When crossing the boundary of regions I and II, the tangential magnetic field component changes its sign.
The slow shock ($ S_{-}^{\,\uparrow} $),
that does not reverse the tangential field component turns into the reversing slow shock ($ {S_{-}^{\,\downarrow}} $).
The transition solution is realized at the
boundary of the regions, when $ \theta_{2} = 0 $.
Substituting $ B_{y2} = 0 $ into (\ref{GRx12}) gives the corresponding transition solution:
\begin{equation}
    \left\{ \, B_{x} \, \right\} = 0 \, ,
    \quad
    \left\{ \, \rho \, v_{x} \, \right\} = 0 \, ,
    \quad
    \left\{ \, p + \rho \, v^{\,2}_{x}\,
    \right\}
  = \frac{B_{y1}^{\,2}}{8 \pi} \, ,
    \label{GRx22}
\end{equation}
\begin{displaymath}
    B_{x} \left\{ \, v_{y} \, \right\} = - v_{x1} B_{y1} \, ,
    \quad
    \rho \, v_{x}\left\{ \, v_{y} \,
    \right\} = - \frac{1}{4 \pi} \, B_{x} B_{y1} \, .
    \label{GRx24}
\end{displaymath}

\vspace{0.3mm}

\noindent
Eliminating $ \{ v_y \} $ from the last two equations, we find
\begin{equation}
   m^2 = {\rho_1 B_x^{\, 2} } / \, { 4 \pi } =
   m_{\rm off}^{\,2} \, ,
\end{equation}
which was to be proved. This mass flux corresponds to the switchoff shock~($ {S_{\, \rm off}} $):
the tangential field component disappears behind the discontinuity plane. This occurs irrespective of the angle~$ \theta_1 $ and corresponds to
the intersection of the curves at $ m^2 = m_{\rm off}^{\,2} $
in Fig.~\ref{fig1}.
%
%
%
%

The reversal of the tangential field component at the boundary of regions II and III can be a special case of a threedimensional Alfv'en discontinuity~($ {A} $).
Since there is no density jump at the Alfv'en discontinuity, let us substitute $ \{ \rho \} = 0 $
into (\ref{GRx12}):
\begin{equation}
    \left\{ \, \rho \, \right\} =
    \left\{ \, B_{x}\, \right\} =
    \left\{ \, v_{x}\, \right\} = 0 \, ,
    \quad
    \left\{ \, p + \frac{ B_{y}^{\,2} }{ 8 \pi } \,
    \right\}
  = 0 \, ,
    \label{GRx27}
\end{equation}
\begin{displaymath}
    B_{x} \left\{ \, v_{y} \, \right\}
  = v_{x} \left\{ \, B_{y} \, \right\} \, ,
    \quad
    \rho \, v_{x} \left\{ \,
    v_{y} \,
    \right\}
  = \frac{1}{ 4 \pi } \, B_{x} \left\{\, B_{y} \, \right\} \, .
    \label{GRx29}
\end{displaymath}

\vspace{0.3mm}

\noindent
If $ \{ B_y \} = 0 $, then all quantities are continuous and there is no discontinuity. Let $ \{ B_y \} \ne 0 $.
Eliminating the ratio $ \{ B_y \} / \{ v_y \} $ from the last two equations, we obtain
\begin{equation}
   m^{\,2} = {\rho B_x^{\, 2} } / \, { 4 \pi }
   = m_{_{\rm A}}^{\,2} \, .
   \label{mAlv}
\end{equation}
When substituting $ \{ \rho \} = 0 $ into
(\ref{GRx1})--(\ref{GRx7}) we find the
boundary conditions at the Alfv'en discontinuity:
\begin{equation}
    \left\{ \, \rho \, \right\} =
    \left\{ \, B_{x} \, \right\} =
    \left\{ \, v_{x} \, \right\} = 0 \, ,
    \quad
    \left\{ \, p +
    \frac{ B_{y}^{\,2 } + B_{z}^{\,2} }{ 8 \, \pi} \,
    \right\}  = 0 \, ,
    \label{GRx34}
\end{equation}
\begin{displaymath}
    B_{x} \left\{ \, v_{y} \, \right\}
  = v_{x} \left\{ \, B_{y} \, \right\} \, ,
    \quad
    \rho \, v_{x} \, \left\{ \, v_{y} \, \right\}
  = \frac{1}{4 \pi} \, B_{x} \left\{ \, B_{y} \, \right\}\, ,
\end{displaymath}
\begin{displaymath}
    B_{x} \left\{ \, v_{z} \, \right\}
  = v_{x} \left\{ \, B_{z} \, \right\} \, ,
    \quad
    \rho \, v_{x} \, \left\{ \, v_{z} \, \right\}
  = \frac{1}{4 \pi} \, B_{x} \left\{ \, B_{z} \, \right\} \, .
    \label{GRx36}
\end{displaymath}

\vspace{0.3mm}

\noindent
Comparison of  (\ref{GRx27}) and (\ref{GRx34}) shows that the boundary conditions (\ref{GRx27}) describe the transition discontinuity between the slow shock in the limit
$ \{ \rho \} \rightarrow 0 $ and the Alfv'en flow at
$ v_z = 0 $ and
$ B_z= 0 $.
The discontinuity is a special case of the Alfv'en discontinuity that reverses the tangential magnetic field component. Trans-Alfv'en discontinuities reverse and enhance the tangential field component. They occupy region III and are adjacent to the Alfv'en mass flux (\ref{mAlv}) on the right. The conditions for the transition to the Alfv'en discontinuity are identical to (\ref{GRx27}).
%
%

There can be no flow near the boundary of regions
III and IV in some range of mass  fluxes. For this rea
son, the transition between the trans-Alfv'en and fast
shock is forbidden. The range narrows as the initial
inclination angle of the magnetic field decreases to
$ \theta_1 = 0 $
(Fig. \ref{fig1}).
%
%
The strongest fast shock takes place at the
minimum possible mass flux admissible by condition (\ref{GRx11.1}).
As the mass flux increases, the tangent of the field
inclination angle behind the discontinuity plane
decreases, asymptotically approaching
$ {\rm tan} \, \theta_2 =( {\rho_2 / \rho_1} ) \,
  {\rm tan} \, \theta_1 $.
%
%

%
%
\section{THREEDIMENSIONAL DISCONTINUOUS FLOWS}

\begin{figure} 
    \centerline{\includegraphics[width=150mm]{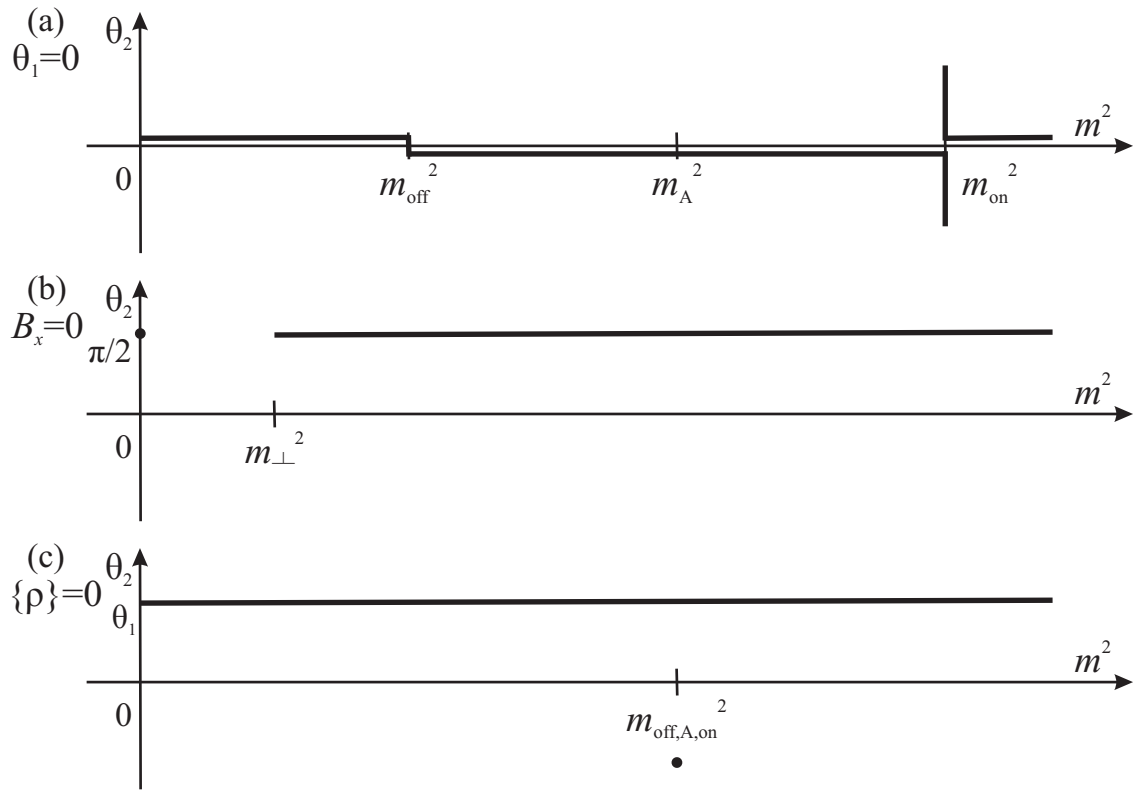}}
    \vspace*{3cm}
\caption{\rm Schematic behavior of the dependence $ \theta_2 \, (m^2) $
       for $ \theta_1 = 0 $~(a), $ B_x = 0$~(b),
       $ \left\{ \rho \right\} = 0 $~(c).}
    \label{fig2}
\end{figure}

Varying $ \rho_1 $, $ \{ \rho \} $ and $ B_x $
leads to contraction or extension of the curves presented in Fig.~\ref{fig1},
%
%
along the coordinate axes without any change of their overall
structure. For zero $ \theta_1 $, $ B_x $ and $ \{ \rho \} $ the behavior of the dependence $ \theta_2 ( m^{2} ) $ is shown in Fig.~\ref{fig2}.
%
%
In view of (\ref{GRx9.2}) when
$ \theta_1 \rightarrow 0 $ the angle $ \theta_2 $
also approaches zero at
almost all values of $ m^2 $
(the case of 
$ m^2 = m_{\rm on}^{\,2} $
will be
considered separately). If $ B_{y1} = 0 $, then $ B_{y2} = 0 $
(Fig.~\ref{fig2}{\it a}).
%
%
In this case, the boundary conditions for twodimensional discontinuities
(\ref{GRx12}) take the form
\begin{equation}
    \left\{ \, \rho \, v_{x} \, \right\} = 0 \, ,
    \quad
    \left\{ \, p + \rho \, v^{2}_{x} \,
    \right\}
  = 0 \, ,
    \label{GRx39}
\end{equation}
\begin{displaymath}
    \left\{ \, B_{x} \, \right\} = 0 \, ,
    \quad
    \left\{ \, v_{y} \, \right\} = 0 \, ,
    \label{GRx41}
\end{displaymath}

\vspace{0.3mm}

\noindent
which corresponds to an ordinary hydrodynamic
shock that propagates according to the conditions $ B_{y1} = 0 $ and $ B_{y2} = 0 $
along the magnetic field. System (\ref{GRx39}) is the transition solution between the oblique shocks in the limit
$ \theta_1 \rightarrow 0 $ and the parallel shock~($ {S_\parallel} $).
%
%

As the angle $ \theta_1 $ decreases, the discontinuity between the admissible mass fluxes for fast and trans-Alfv'en shocks will also decrease. Conditions (\ref{GRx12}) in this case give
\begin{equation}
    \left\{ \, B_{x} \, \right\} = 0 \, ,
    \quad
    \left\{ \, \rho \, v_{x} \, \right\} = 0 \, ,
    \quad
    \left\{ \, p + \rho \, v^{\,2}_{x} \,
    \right\}
  = - \frac{ B_{y2}^{\,2} }{ 8 \pi } \, ,
    \label{GRx43}
\end{equation}
\begin{displaymath}
    B_{x} \left\{ \, v_{y} \, \right\} = v_{x2} B_{y2} \, ,
    \quad
    \rho \, v_{x} \left\{ \, v_{y} \,
    \right\}
    = \frac{1}{ 4 \pi } \, B_{x} B_{y2} \, .
    \label{GRx45}
\end{displaymath}

\vspace{0.3mm}

\noindent
From the simultaneous solution of the last two equations, we have
\begin{equation}
    m^2
  = { \rho_2 B_x^{\, 2} / \, 4 \pi } =
    m_{\rm on}^{\,2} .
\end{equation}
For this mass flux, Eq.  (\ref{GRx9.2})
has no unique solution. A
nonzero $ \theta_1 $ can correspond to zero $ \theta_2 $.
A tangential
magnetic field component appears behind the shock
front, corresponding to the switchon shock~($ { S_{\, \rm{on}} } $).
It is indicated in Fig. \ref{fig2}{\it a}
%
%
by the vertical segment at
$ m^2 = m^2_{\, \rm{on}} $.
The switchon shock can act as the transition one
for the trans-Alfv'en and fast shocks in the limit
$ \theta_1 \rightarrow 0 $, but this, of course, requires that
$ m^2 \rightarrow m^2_{\, \rm{on}} $.
Other
wise, there will be the transition to the parallel shock according to (\ref{GRx39}).
%
%

To establish the form of the transition solution between the parallel shock and the contact discontinuity, we will set 
$ v_x = 0 $ in (\ref{GRx39}). We then have
\begin{equation}
    \left\{ \, B_{x} \, \right\} =
    \left\{ \, v_{y} \, \right\} =
    \left\{ \, p \, \right\} = 0 \, ,
    \label{GRx49}
\end{equation}

\vspace{0.3mm}

\noindent
This system of equations corresponds to the contact discontinuity
(\ref{GRx17}), orthogonal to the magnetic field lines. It describes the transition discontinuity between the parallel shock in the limit $ v_x \rightarrow 0 $ and the contact
discontinuity. Such a transition takes place at $ m^2 = 0 $ in Fig.~\ref{fig2}{\it a}.
%
%
%
%

When $ B_x = 0 $ (Fig.~\ref{fig2}{\it b}),
%
%
 $ m^2_{\, \rm{on}} $ becomes zero and all nonzero mass fluxes are in region IV in Fig.~\ref{fig1}.
%
%
To find the boundary conditions corresponding to them, let us substitute $ B_x = 0 $ into (\ref{GRx12}).
We obtain
\begin{equation}
    \left\{ \, \rho \, v_{x} \, \right\} = 0 \, , \quad
    \left\{ \,
    p + \rho \, v^{\,2}_{x} + \frac{ B_{y}^{\,2} }{ 8 \pi }
    \right\}
  = 0 \, ,
    \label{GRx51}
\end{equation}
\begin{displaymath}
    \left\{ \, v_{x} B_{y} \, \right\} = 0 \, , \quad
    \left\{ \, v_{y} \, \right\} = 0 \, .
    \label{GRx53}
\end{displaymath}

\vspace{0.3mm}

\noindent
These conditions characterize a compression shock propagating perpendicularly to the magnetic field. In the general case of a perpendicular shock~($ {S_\bot} $) we will find the boundary conditions by substituting
$ B_x = 0 $ into (\ref{GRx1})--(\ref{GRx7}):
\begin{displaymath}
    \left\{ \, \rho \, v_{x} \, \right\} = 0 \, ,
    \quad
    \left\{ \,
    p + \rho \, v^{\,2}_{x}
      + \frac{B_{y}^{\,2} + B_{z}^{\,2}}{8 \pi} \,
    \right\}
  = 0 \, ,
    \label{GRx59}
\end{displaymath}
\begin{equation}
    \left\{ \, v_{x} B_{y} \, \right\} = 0 \, ,
    \quad
    \left\{ \, v_{y} \, \right\} = 0 \, ,
    \label{GRx58}
\end{equation}
\begin{displaymath}
    \left\{ \, v_{x} B_{z} \, \right\} = 0 \, ,
    \quad
    \left\{ \, v_{z} \, \right\} = 0 \, .
\end{displaymath}

\vspace{0.3mm}

\noindent
Equations (\ref{GRx51}) are then the boundary conditions for
the transition discontinuity between the fast shock in the limit $ B_x \rightarrow 0 $ and the perpendicular shock with
the magnetic field directed along the $ y $
axis. This transition can take place only at mass fluxes that satisfy
inequality (\ref{GRx11}), which takes the form $ m^{\,2} > m^{\,2}_\bot $ at
$ B_x = 0 $
(Fig.~\ref{fig2}{\it b}).
%
%
%
%

To determine the boundary conditions for the dis
continuity at
$ m^2 = 0 $ (Fig.~\ref{fig2}{\it c}),
%
%
let us substitute $ B_x = 0 $ and
$ v_x = 0 $ into (\ref{GRx1})--(\ref{GRx7}).
In this case, the magnetic field and the velocity field are parallel to the discontinuity surface and can undergo arbitrary jumps in magnitude and direction, while the jump in pressure is related to the jump in magnetic field by the condition
\begin{equation}
    \left\{ \,
    p + \frac{ B_{y}^{\,2} + B_{z}^{\,2} }{ 8 \pi} \,
    \right\}
  = 0 \, .
    \label{GRx63}
\end{equation}

\vspace{0.3mm}

\noindent
This corresponds to the tangential discontinuity~($ {T} $).
The contact discontinuity, the slow shock, and the Alfv'en discontinuity can pass to it in the limit $ B_x \rightarrow 0 $ under certain conditions. Let us find the corresponding transition solutions. First, let us substitute
 $ B_x = 0 $ into the boundary conditions for the contact discontinuity (\ref{GRx17}), We will obtain the transition solution
\begin{equation}
    \left\{ \, B_{y} \, \right\} =
    \left\{ \, v_{y} \, \right\} =
    \left\{ \, p \, \right\} = 0 \, .
    \label{GRx62}
\end{equation}

\vspace{0.3mm}

\noindent
It describes the tangential discontinuity (\ref{GRx63}) for zero field component $ B_z $ in the absence of jumps
$ \{ v_y \} $ and $ \{ B_y \} $.
%
%
%
Second, the conditions for oblique shocks
(\ref{GRx12}) at $ B_x = 0 $ and $ v_x = 0 $ are the transition solution
\begin{equation}
    \left\{ \, p + \frac{ B_{y}^{\,2} }{ 8 \pi } \,
    \right\}
  = 0 \, ,
    \label{GRx61}
\end{equation}

\vspace{0.3mm}

\noindent
that corresponds to the plane tangential discontinuity (\ref{GRx63}) at $ B_z = 0 $.
Finally, the boundary condition for the Alfv'en discontinuity
(\ref{GRx34}) after the substitution of $ B_x = 0 $ gives the transition solution
\begin{equation}
    \left\{ \, \rho \, \right\} = 0 \, , \quad
    \left\{ \, p + \frac{ B_{y}^{\,2} + B_{z}^{\,2} }{ 8 \pi } \,
    \right\}
  = 0 \, ,
    \label{GRx60}
\end{equation}

\vspace{0.3mm}

\noindent
that describes the tangential discontinuity (\ref{GRx63}), with
out any jump in density $ \rho $.
%
%

The angle $ \theta_2 $ for the strongest trans-Alfv'en shock is
defined by Eq. (\ref{max3}).
In view of (\ref{max3}), $ \theta_2 \rightarrow - \theta_1 $ when $ B_x \rightarrow 0 $.
Thus, the trans-Alfv'en shocks degenerate into a special case of the Alfv'en discontinuity as the magnetic flux decreases. Of course, the density jump can also be set equal to zero for any type of flow. In this
case, all parameters
$ m^2_{\, \rm{off}} $, $ m_{_{ \rm{A} }}^{\, 2} $ and
$ m^2_{\, \rm{on}} $
will be equal to the flux (\ref{mAlv}), at which the Alfv'en discontinuity (\ref{GRx27}).
takes place. In Fig. \ref{fig2}{\it c},
it is denoted by $ m^2_{\, \rm{off,A,on}} $.
At other mass fluxes, the differences in plasma character
istics on different sides of the discontinuity will disappear; the discontinuity will be absent as such.
%
%

\vspace{0.3mm}

\begin{figure} 
    \centerline{\includegraphics[width=150mm]{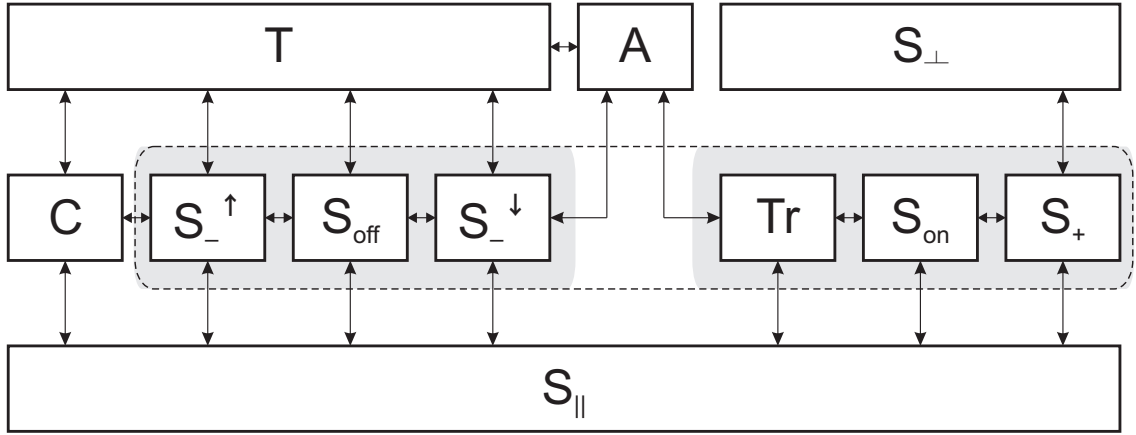}}
    \vspace*{3cm}
\caption{\rm Scheme of continuous transitions between MHD discontinuities. The dashed line encircles the set of discontinuities corresponding to the block of oblique shocks in Syrovatskii's scheme
       The 	
pouring inside the contour highlights the ``slow'' (left) and ``fast'' (right) components of the scheme proposed in
       \cite{Somov-94}.}
    \label{fig3}
\end{figure}

Let us combine the properties of discontinuous solutions systematized above into the scheme of permitted transitions shown in Fig.~\ref{fig3}.
%
%
Here, the twodimensional discontinuities are located in the middle row in order of increasing mass flux and the threedimensional discontinuities are located in the upper row. The onedimensional parallel shock~($ S_{\, \parallel} $).
occupies the lower row. The individual elements are grouped for the convenience of comparing the generalized scheme of transitions with those proposed previously. Syrovatskii's scheme
\cite{Syrovatskii-56}
is consistent with Fig. \ref{fig3},
%
%
if we combine the elements~$ {S_{-}^{\,\downarrow}} $,
$ {S_{\, \rm{off}}} $, $ {S_{-}^{\,\uparrow}} $, $ {Tr} $,
$ {S_{\, \rm{on}}} $ and
$ {S_{+}} $ into one block of ``oblique shocks''~($ {S} $), while omitting the question of whether any transitions inside the block are possible and disregard the contact discontinuity~($ {C} $)and the parallel shock~($ {S_{\, \parallel}} $).
The scheme proposed in
\cite{Somov-94},
%
includes the parallel shock~($ {S_{\, \parallel}} $)
and the separation of oblique shocks into the slow one ($ {S_{-}} $), corresponding to condition (\ref{GRx10.1}), and the fast one~($ {S_+} $),
corresponding to condition
(\ref{GRx11.1}).
It is quite obvious that the scheme of transitions we propose is a proper and natural generalization of the two previous schemes. Our scheme contains not only evolutionary types of discontinuities but also nonevolutionary ones: the switchon, Alfv'en, and trans-Alfv'en shocks.
%
%

\vspace{0.5mm}

\section{THE JUMP IN INTERNAL ENERGY}
\label{sec5}

To determine the plasma heating efficiency, let us turn to the boundary condition (\ref{GRx8}), which is the
energy conservation law. Using (\ref{GRx2}), we will find the jump in internal energy from (\ref{GRx8}) 
\begin{equation}
      \left\{ \, \epsilon \, \right\}
  = - \left\{ \frac{ v^{\,2} }{2} \right\}
    - \frac{1}{m} \left\{ \, v_{x} p \, \right\}
    - \frac{1}{ 4 \pi m }
      \left\{ \, B^{\,2} v_{x}
    - ( \, {\bf v} \cdot {\bf B} \, ) \, B_{x} \, \right\} .
      \label{GRx65}
\end{equation}
Using the mean velocities $ \tilde v_{x} $,
$ \tilde v_{y} $ and $ \tilde v_{z} $, we will write the first term as
\begin{displaymath}
    - \left\{ \frac{v^{\,2}}{2} \right\}
  = - \tilde v_{x} \left\{ v_{x} \right\}
    - \tilde v_{y} \left\{ v_{y} \right\}
    - \tilde v_{z} \left\{ v_{z} \right\}.
      \label{GRx66}
\end{displaymath}
We will express the jumps in tangential velocity components in terms of the jumps in tangential magnetic field components using Eqs. (\ref{GRx1}), (\ref{GRx2}),
(\ref{GRx5}) and (\ref{GRx6}) as
\begin{displaymath}
    \left\{ v_{y} \right\}
  = \frac{ B_x }{ 4 \pi m } \left\{ B_{y} \right\} , \quad
    \left\{ v_{z} \right\}
  = \frac{ B_x }{ 4 \pi m } \left\{ B_{z} \right\} .
    \label{GRx67}
\end{displaymath}
Now,the first term on the right side of Eq. (\ref{GRx65})
appears as
\begin{equation}
    - \left\{ \frac{v^{\,2}}{2} \right\}
  = - \tilde v_{x} \left\{ v_{x} \right\}
    - \frac{ \tilde v_{y} B_{x} }{ 4 \pi m} \, \{ B_{y} \}
    - \frac{ \tilde v_{z} B_{x} }{ 4 \pi m} \, \{ B_{z} \} .
    \label{GRx68}
\end{equation}
Similarly, the entire right side of Eq. (\ref{GRx65}) can be expressed in terms of the jumps in normal velocity components and tangential magnetic field components:
\begin{displaymath}
      \left\{ \, \epsilon \, \right\}
  = - \frac{ \tilde p }{ m } \, \{ v_{x} \}
    + \frac{ \tilde v_{x} \tilde B_{y}
           - \tilde v_{y} B_{x} }{ 4 \pi m } \, \{ B_{y} \}
    - \frac{ v_x B_y - v_y B_x }{ 4 \pi m } \, \{ B_y \} \, +
\end{displaymath}
\begin{displaymath}
  + \, \frac{ \tilde v_{x} \tilde B_{z}
         - \tilde v_{z} B_{x} }{ 4 \pi m } \, \{ B_{z} \}
  - \frac{ v_x B_z - v_z B_x }{ 4 \pi m } \, \{ B_z \} .
    \label{GRx75}
\end{displaymath}
This equation can be simplified if we expand the means appearing in it and take
$ - \{v_x\} / m = - \{r\} $.
outside the brackets. We obtain
\begin{equation}
      \left\{ \, \epsilon \, \right\}
  = - \{ r \}
      \left( \tilde p
    + \frac{ \{ B_y \}^2 + \{ B_z \}^2 }{ 16 \pi }
      \right) .
    \label{GRx77}
\end{equation}
%
For twodimensional discontinuities, it takes quite a
simple form,
\begin{equation}
    \left\{ \, \epsilon \, \right\}
  = - \{ r \}
    \left( \tilde p + \frac{ \{ B_y \}^2 }{ 16 \pi }
    \right) .
    \label{GRx78}
\end{equation}
%
%

\begin{figure} 
    \centerline{\includegraphics[width=150mm]{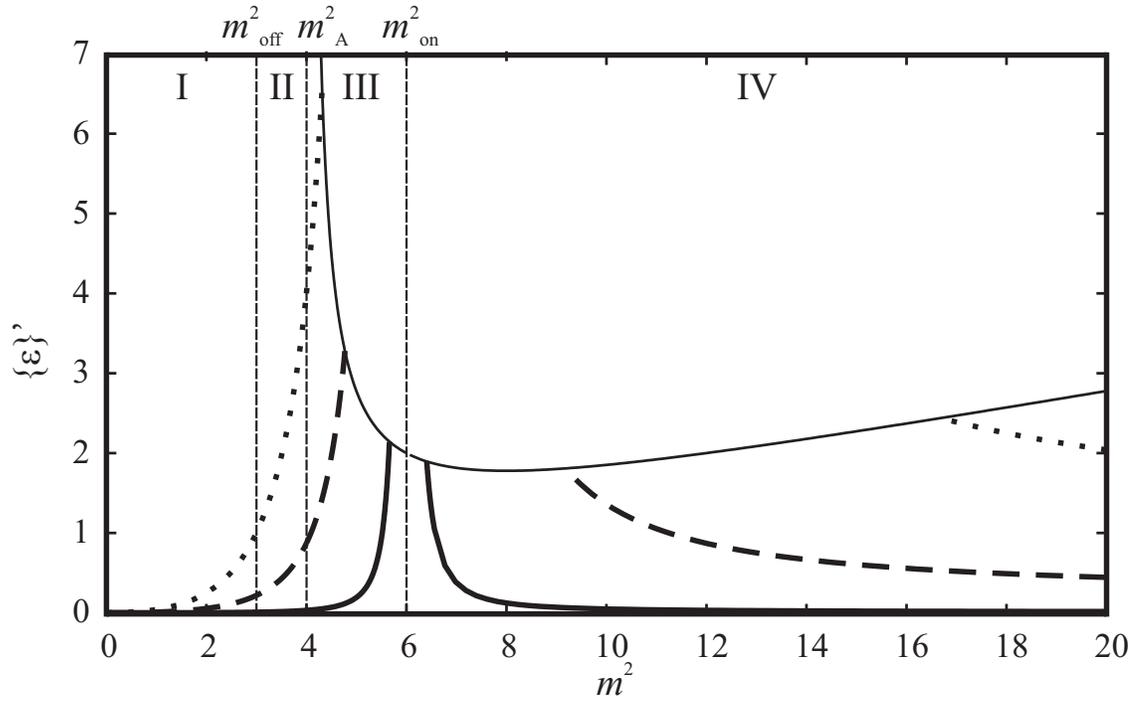}}
    \vspace*{3cm}
\caption{\rm Jump in internal energy $ \epsilon $ versus mass flux for various angles
       $ \theta_{1} $: $ 5^{\circ} $ (solid lines),
       $ 25^{\circ} $(dashed lines), and $ 45^{\circ} $ (dotted lines).}
    \label{fig4}
\end{figure}

Equation (\ref{GRx77}) allows definitive conclusions regarding the change in plasma internal energy when crossing the discontinuity surface to be reached. First, the internal energy increases, because, according to Zemplen's theorem, $ - \{ r \} > 0 $ and $ \tilde p $ and $ \{ B_y \}^2 $ are positive. Second, the change in internal energy consists of two parts: the thermodynamic and magnetic ones. The latter depends on the magnetic field configuration and, hence, on the type of discontinuity. Let us express the tangential magnetic field components in Eq.
 (\ref{GRx78}) in terms of the corresponding inclination angles:
\begin{displaymath}
    \left\{ \, \epsilon \, \right\}
  = - \{ r \} \, \tilde p
    - \{ r \} \, \frac{B_x^{\, 2}}{ 16 \pi }
    \left( {\rm tan} \, \theta_2
         - {\rm tan} \, \theta_1 \right)^{\, 2} .
\end{displaymath}
Then, we will take the thermodynamic part of the heating independent of the type of discontinuity as the zero point and will measure the jump in internal
energy itself in units of $ - \{ r \} {B_x^{\,2}}/\,{ 16 \pi } $.
For this purpose, let us make the substitution
\begin{displaymath}
    \left\{ \, \epsilon \, \right\}^{\, \prime}
  = -  \frac{ 16 \pi }{ \{ r \} B_x^{\, 2} }
    \left( \left\{ \, \epsilon \, \right\} + \{ r \} \tilde p
    \right)\, .
\end{displaymath}
We will obtain the equation
\begin{equation}
    \left\{ \, \epsilon \, \right\}^{\, \prime}
  = \left( {\rm tan} \, \theta_2 - {\rm tan} \, \theta_1
    \right)^{\, 2} \, .
    \label{GRx79}
\end{equation}
For the discontinuities presented in Fig. \ref{fig1},
%
%
the dependence of the jump in internal energy on mass flux was calculated using Eq. (\ref{GRx79}) and is shown in Fig. \ref{fig4}.
%
%
The curve describing the jumps in internal energy at the strongest trans-Alfv'en and fast shocks is
\begin{equation}
    \left\{ \, \epsilon \, \right\}^{\, \prime} =
    \frac{ \left(m^{\, 2} / \, m_{\rm off}^{\, 2}
    - \, m^{\, 2} / \, m_{\rm on}^{\, 2} \right)^2 }
         { m^{\,2} / \,m_{_{\rm A}}^{\,2} - 1 } \, .
\label{GRx80}
\end{equation}
It is indicated by the thin line in Fig. \ref{fig4}.
%
%

We see that the maximum jump in internal energy
is produced by the strongest trans-Alfv'en shock, with its magnitude increasing rapidly with angle $ \theta_1 $.
The efficiency of plasma heating by other types of discontinuities depends on specific conditions. For example, the heating by slow shocks can be both lower than the heating by fast shocks at smaller $ \theta_1 $,
and higher at larger $ \theta_1 $.
In any case, the heating depends on the shock strength. The larger the change in magnetic energy density, the higher the temperatures to which the plasma will be heated.

\vspace{1mm}
%
%

\section{CONCLUSIONS}
\label{sec6}

We considered the boundary conditions for the equations of ideal MHD at the surfaces of discontinuities of various types and established the specific form of transition solutions for possible pairs of discontinuous flows. Based on them, we constructed a generalized scheme of permitted transitions that contains not only evolutionary types of discontinuities but also nonevolutionary ones: the switchon, Alfv'en, and trans-Alfv'en shocks. When interpreting the results of our numerical integration of the MHD equations, this allows us to see the regions that require a more careful calculation and a further study. In particular, in the problem of magnetic reconnection in current layers, the discontinuous flows attached to the ends of the current layer, to the reverse currents, are such regions
\cite{Somov-13}.
We derived the equation describing the change in plasma internal energy when passing through the discontinuity. We established its dependence on the type of discontinuity. The larger the jumps in plasma density and magnetic energy density at the discontinuity, the stronger the heating. The formation of nonevolutionary trans-Alfv'en shocks also contributes to the plasma heating. In the phenomena driven by magnetic reconnection, for example, in solar flares, this can probably explain the observed distribution of the highest plasma temperatures in the solar corona
\cite{Sui-03,Somov-13}.

\vspace{0.5mm}

%
%
\section{ACKNOWLEDGMENTS}
This work was supported by the Russian Foundation for Basic Research (project no. 110200843a).
%

%
%

\vspace{1cm}
Translated by V. Astakhov
\clearpage


%
%

%
%
\clearpage

%
%
\clearpage

%
%
\clearpage

%
%
\end{document}